\documentclass[10pt,twocolumn,twoside,conference]{IEEEtran}

\usepackage{graphicx}
\usepackage{cite}
\usepackage[cmex10]{amsmath}
\usepackage{amssymb}

\usepackage{tikz}
\usepackage{pgfplots}
\pgfplotsset{compat=1.3}
\usetikzlibrary{arrows,snakes,shapes}

\DeclareMathAlphabet{\mathbit}{OML}{cmr}{bx}{it}

\DeclareMathOperator{\E}{E}
\DeclareMathOperator{\T}{T}
\DeclareMathOperator{\Tr}{Tr}

\DeclareMathOperator{\Probability}{Pr}

\renewcommand\arcsin[1]{\operatorname{arcsin}\left(#1\right)}

\DeclareMathOperator{\fieldR}{\mathbb{R}}

\newcommand\vech[1]{\operatorname{vech}\left(#1\right)}

\newcommand{\sign}[1]{\Sign{\left\{#1\right\}}}

\newcommand{\ve}[1]{\boldsymbol{#1}}

\newcommand{\exdi}[2]{\E_{#1} \left[#2\right]}
\newcommand{\ex}[1]{\E \left[#1\right]}

\renewcommand{\exp}[1]{\operatorname{exp}\left(#1\right)}

\newcommand{\tr}[1]{\Tr \left(#1\right)}
\newcommand{\Prob}[1]{\Probability\left\{#1\right\}}

\newcommand\Sign{\operatorname{sign}}

\title{DOA Parameter Estimation with $1$-bit Quantization\\Bounds, Methods and the Exponential Replacement}

\author{
\IEEEauthorblockN{Manuel~Stein, Kurt Barb\'{e}\IEEEauthorrefmark{1} and Josef A. Nossek}
\IEEEauthorblockA{Institute for Circuit Theory and Signal Processing (NWS), Technische Universit\"at M\"unchen, Germany} 
\IEEEauthorblockA{\IEEEauthorrefmark{1}Research Team Stochastics (STOX), Dept. Mathematics (DWIS), Vrije Universiteit Brussel, Belgium}    
Email: manuel.stein@tum.de, kurt.barbe@vub.ac.be, josef.a.nossek@tum.de}

\begin{document}
\maketitle
\begin{abstract}
While $1$-bit analog-to-digital conversion (ADC) allows to significantly reduce the analog complexity of wireless receive systems, using the exact likelihood function of the hard-limiting system model in order to obtain efficient algorithms in the digital domain can make $1$-bit signal processing challenging. If the signal model before the quantizer consists of correlated Gaussian random variables, the tail probability for a multivariate Gaussian distribution with $N$ dimensions (general orthant probability) is required in order to formulate the likelihood function of the quantizer output. As a closed-form expression for the general orthant probability is an open mathematical problem, formulation of efficient processing methods for correlated and quantized data and an analytical performance assessment have, despite their high practical relevance, only found limited attention in the literature on quantized estimation theory. Here we review the approach of replacing the original system model by an equivalent distribution within the exponential family. For $1$-bit signal processing, this allows to circumvent calculation of the general orthant probability and gives access to a conservative approximation of the receive likelihood. For the application of blind direction-of-arrival (DOA) parameter estimation with an array of $K$ sensors, each performing $1$-bit quantization, we demonstrate how the exponential replacement enables to formulate a pessimistic version of the Cram\'er-Rao lower bound (CRLB) and to derive an asymptotically achieving conservative maximum-likelihood estimator (CMLE). The $1$-bit DOA performance analysis based on the pessimistic CRLB points out that a low-complexity radio front-end design with $1$-bit ADC is in particular suitable for blind wireless DOA estimation with a large number of array elements operating in the medium SNR regime.
\end{abstract}
\section{Introduction}
Concerning hardware complexity and energy consumption of signal processing systems, the circuit forming the analog-to-digital converter (ADC) at the receiver has been identified as a bottleneck \cite{Walden99}. While a high number of bits $b$ allows accurate representation of the analog signals in the digital domain and therefore high processing performance, the power dissipation and production cost of the ADC device scales exponentially $\mathcal{O}(2^b)$ with $b$. An interesting approach is to reduce the resolution of the ADC and resort to a simple device without feedback. In the extreme case the continuous analog waveform at each sensor is directly converted into a binary representation by a hard-limiter. The circuit for such an ADC can be realized by a single comparator, making the ADC highly efficient with respect to its hardware and energy requirements. Further, the binary structure of the resulting receive signal allows to perform basic signal processing operations in the digital domain by efficient $1$-bit arithmetics \cite{Host00}. Therefore, $1$-bit ADC also provides a beneficial effect on the complexity of the digital processing unit of the receive system. 

Nevertheless, a serious drawback of $1$-bit ADC is that a highly nonlinear and noninvertible operation is performed on the original analog receive signal. In comparison to an ideal receive system with $\infty$-bit ADC resolution this is associated with a significant performance loss. It is well understood that for certain problems, i.e., location parameter estimation with uncorrelated noise, the performance gap between a symmetric hard-limiting $1$-bit system and an ideal receiver with infinite ADC resolution is moderate ($2/\pi$ or $-1.96$ dB) in the low signal-to-noise ratio (SNR) regime \cite{Vleck66} \cite{Stein15}. However, important technologies like wireless communication usually take place in the medium SNR regime where the quantization loss is more pronounced. In particular for such scenarios we have identified that introducing redundancy by modification of the analog processing prior to the ADC, allows to recover large portions of the $1$-bit loss \cite{SteinWCL15}. To analyze this effect, we have developed a compact lower bound for the Fisher information measure \cite{Stein14SPL} and recently generalized it, in order to obtain better bounding accuracy for advanced signal parameter estimation problems \cite{SteinTSP16}. Further, we have shown that the underlying methodology of exponential replacement leads to an approximate formulation of the likelihood function which can be used to formulate point estimation algorithms that achieve the conservative inference capability of the nonlinear stochastic system in a consistent way\cite{SteinTSP16}. While in \cite{SteinWCL15} we focused on parametric location parameter estimation of a hard-limited multivariate Gaussian variable with fixed covariance matrix, here we center the discussion around the estimation of a parameter modulating the covariance matrix of a zero-mean multivariate Gaussian variable after $1$-bit quantization. This kind of estimation problem arises in wireless applications if the direction-of-arrival (DOA) parameter of a transmit signal with unknown structure impinging on a receive array is to be determined. DOA parameter estimation forms a specific application of blind covariance-based estimation and plays a key role for technologies like wireless multi-user communication, spectrum monitoring, jammer localization and interference mitigation for satellite-based radio systems.
\section{Motivation} \label{sec:motivation}
In order to clearly outline the estimation theoretic motivation behind the presented work, we review efficient covariance-based estimation with infinite ADC resolution and outline the challenges arising when treating a receiver with $1$-bit ADC.
\subsection{Ideal Receiver ($\infty$-bit ADC)}
Consider a digital receive signal $\ve{y}\in\fieldR^{M}$ which is well represented by a multivariate Gaussian random variable with probability density function
\begin{align}\label{multi:gauss:model}
p(\ve{y};\theta)=\frac{1}{(2\pi)^{\frac{N}{2}} \sqrt{\det \ve{\Sigma}_{\ve{y}}(\theta) }} \exp{- \frac{1}{2} \ve{y}^{\T} \ve{\Sigma}^{-1}_{\ve{y}}(\theta)\ve{y}}
\end{align}
with a single parameter $\theta\in\Theta$ modulating the covariance
\begin{align}
\ve{\Sigma}_{\ve{y}}(\theta)=\exdi{\ve{y};\theta}{\ve{y}\ve{y}^{\T}}
\end{align}
and vanishing mean
\begin{align}
\exdi{\ve{y};\theta}{\ve{y}}=\ve{0}, \quad\quad \forall \theta.
\end{align}
Given $N$ independent data snapshots
\begin{align}
\ve{Y}=\begin{bmatrix} \ve{y}_1 &\ve{y}_2 &\ldots &\ve{y}_N\end{bmatrix} \in \fieldR^{M\times N},
\end{align}
the optimum asymptotically unbiased estimator $\hat{\theta}(\ve{Y})$ is obtained by maximizing the likelihood \cite{Kay93}
\begin{align}\label{def:mle}
\hat{\theta}(\ve{Y}) &= \arg \max_{\theta\in\Theta} \ln p(\ve{Y};\theta)\notag\\
&= \arg \max_{\theta\in\Theta} \sum_{n=1}^{N} \ln p(\ve{y}_n;\theta)\notag\\
&= \arg \min_{\theta\in\Theta} \ln \big(\det \ve{\Sigma}_{\ve{y}}(\theta)\big) + \Tr{\big(\ve{\bar{\Sigma}}_{\ve{y}}(\ve{Y})\ve{\Sigma}_{\ve{y}}^{-1}(\theta) \big)},
\end{align}
where the receive covariance is given by
\begin{align}
\ve{\bar{\Sigma}}_{\ve{y}}(\ve{Y}) = \frac{1}{N} \sum_{n=1}^{N} \ve{y}_n\ve{y}_n^{\T}.
\end{align}
As the maximum-likelihood estimator is consistent and efficient, it is possible to characterize its asymptotic performance in an analytical way through the Cram\'er-Rao lower bound \cite{Kay93}
\begin{align}
\exdi{\ve{Y};\theta}{\big(\theta-\hat{\theta}(\ve{Y})\big)^{2}} \geq \frac{1}{NF_{\ve{y}}(\theta)},
\end{align}
where the Fisher information is defined 
\begin{align}
F_{\ve{y}}(\theta)=\int_{\ve{\mathcal{Y}}} \bigg( \frac{\partial  \ln p(\ve{y};\theta)}{\partial \theta}  \bigg)^{2} {\rm d} \ve{y}.
\end{align}
For the multivariate Gaussian model \eqref{multi:gauss:model}, we obtain\cite[p. 47]{Kay93}
\begin{align}
F_{\ve{y}}(\theta)&=\frac{1}{2}\tr{ \ve{\Sigma}^{-1}_{\ve{y}}(\theta) \frac{\partial \ve{\Sigma}_{\ve{y}}(\theta)}{\partial \theta} \ve{\Sigma}^{-1}_{\ve{y}}(\theta) \frac{\partial \ve{\Sigma}_{\ve{y}}(\theta)}{\partial \theta} }.
\end{align}
\subsection{Low-Complexity Receiver ($1$-bit ADC)}
The situation changes fundamentally if a nonlinear transformation
\begin{align}
\ve{z}=\ve{f}{(\ve{y})}
\end{align}
is involved. Then the derivation of an exact representation of the likelihood $p(\ve{z};\theta)$, and therefore processing the data with the maximum-likelihood approach \eqref{def:mle}, can become difficult. If we assume $1$-bit ADC and model the converters by an element-wise hard-limiter which discards the amplitude information
\begin{align}
\ve{z}=\operatorname{sign}{(\ve{y}}),
\end{align}
where the element-wise signum function is defined by
\begin{align}
\left[\operatorname{sign}{(\ve{x}})\right]_n=
\begin{cases}
+1& \text{if } x_n \geq 0\\
-1& \text{if } x_n < 0,
\end{cases}
\end{align}
then the likelihood function for one output constellation is found by evaluating the integral
\begin{align}\label{likeli:difficult}
p(\ve{z};\theta)=\int_{\ve{\mathcal{Y}}(\ve{z})} p(\ve{y};\theta) {\rm d}\ve{y}.
\end{align}
Note, that $\ve{\mathcal{Y}}(\ve{z})$ is the subset of $\ve{\mathcal{Y}}$ which is mapped to the output signal $\ve{z}$. Computing such an integral requires the orthant probability of a multivariate Gaussian variable (multivariate version of the Q-function). Unfortunately, a general closed-form expression for the orthant probability is an open mathematical problem. Only for the cases $M\leq4$ solutions are provided  in literature \cite{Kedem80} \cite{Sinn11}. The problem becomes even worse, if one is interested in analytically evaluating the estimation performance of the $1$-bit receive system. The associated Fisher information measure
\begin{align}\label{fisher:difficult}
F_{\ve{z}}(\theta)&=\int_{\ve{\mathcal{Z}}} \bigg( \frac{\partial  \ln p(\ve{z};\theta)}{\partial \theta}  \bigg)^{2} {\rm d} \ve{z}\notag\\
&=\sum_{\ve{\mathcal{Z}}} \bigg( \frac{\partial  \ln p(\ve{z};\theta)}{\partial \theta}  \bigg)^{2}
\end{align}
is computed by summing the squared score function over the discrete support of $\ve{z}$. As $\ve{\mathcal{Z}}$ contains $2^M$ possible receive constellations, direct computation of $F_{\ve{z}}(\theta)$ is prohibitively complex when $M$ is large.
\section{Related Work and Outline}
Due to the outlined problems \eqref{likeli:difficult} and \eqref{fisher:difficult}, the literature on analytic performance bounds and maximum-likelihood algorithms for parametric covariance estimation with $1$-bit quantization is limited. While \cite{Ben48} \cite{Vleck66} are classical references for signal processing with $1$-bit quantizer, more recently \cite{Host00} covers the problem of signal parameter estimation from coarsely quantized data with uncorrelated noise. The work \cite{BarShalom02} is concerned with $1$-bit DOA estimation, but has to restrict the analytical discussion to $K=2$ sensors due to the outlined problem \eqref{fisher:difficult} and resort to empirical methods of high computational complexity for $K>2$. In contrast \cite{Dabeer08} studies location parameter estimation with a multivariate model and dithered $1$-bit sampling while \cite{Jacovitti94} discusses inference of the autocorrelation function from hard-limited Gaussian signals. 

Here we review the approach of exponential replacement which forms the basis for a generalized Fisher information lower bound  \cite{SteinTSP16}. For the application of DOA estimation with $1$-bit quantized data this allows to derive a pessimistic Cram\`{e}r-Rao performance bound and to analyze the achievable estimation accuracy with an arbitrary number of sensors $K$. Further, the exponential replacement enables to state a conservative version of the maximum-likelihood estimator which asymptotically performs equivalent to the presented pessimistic performance bound \cite{SteinTSP16}.
\section{The Exponential Replacement}
Consider an intractable parametric probabilistic model $p(\ve{z};\theta)$. Choose the vector
\begin{align}
\ve{\phi} (\ve{z})  =  \begin{bmatrix} \phi_1(\ve{z}) &\phi_2(\ve{z}) &\ldots &\phi_L(\ve{z}) \end{bmatrix}^{\T}
\end{align}
to be a set of $L$ arbitrary transformations 
\begin{align}
\phi_l(\ve{z}): \fieldR^M \to \fieldR
\end{align}
of the output variable $\ve{z}$ and assume existence and access to the two moments
\begin{align}\label{aux:mean}
\ve{\mu}_{\ve{\phi}}(\theta)=\exdi{\ve{z};\theta}{\ve{\phi}(\ve{z})}
\end{align}
and
\begin{align}\label{aux:covariance}
\ve{R}_{\ve{\phi}}(\theta) &=  \exdi{\ve{z};\theta}{\ve{\phi}(\ve{z})\ve{\phi}^{\T}(\ve{z})} - \ve{\mu}_{\ve{\phi}}(\theta)\ve{\mu}_{\ve{\phi}}^{\T}(\theta).
\end{align}
Then replace the original model $p(\ve{z};\theta)$ by an equivalent model $\tilde{p}(\ve{z};\theta)$ within the exponential family, i.e., a model with sufficient statistics $\ve{\phi} (\ve{z})$ and equivalent moments \eqref{aux:mean} and \eqref{aux:covariance} for which the score function factorizes
\begin{align}\label{approx:score}
\frac{\partial \ln \tilde{p}(\ve{z};\theta)}{\partial \theta} &= \ve{\beta}^{\T}(\theta) \ve{\phi} (\ve{z}) - \alpha(\theta).
\end{align}
After optimizing the weights $\ve{\beta}(\theta)\in\fieldR^{L}$ and $ \alpha(\theta)\in\fieldR$ it can be shown that the Fisher information measure of the original model $p(\ve{z};\theta)$ is in general lower bounded by \cite{SteinTSP16} 
\begin{align}\label{bound:eigenvalue}
F_{\ve{z}}(\theta)\geq\bigg(\frac{\partial \ve{\mu}_{\ve{\phi}}(\theta)}{\partial \theta}\bigg)^{\T}  \ve{R}_{\ve{\phi}}^{-1}(\theta)\frac{\partial \ve{\mu}_{\ve{\phi}}(\theta)}{\partial \theta}.
\end{align}
Further, with $N$ independent output samples
\begin{align}
\ve{Z}=\begin{bmatrix} \ve{z}_1 &\ve{z}_2 &\ldots &\ve{z}_N  \end{bmatrix},
\end{align}
forming the sample mean
\begin{align}
\ve{\tilde{\phi}}=\frac{1}{N}\sum_{n=1}^{N} \ve{\phi} (\ve{z}_n)
\end{align}
and solving the equation
\begin{align}\label{algo:cmle}
\bigg(\frac{\partial \ve{\mu}_{\ve{\phi}}(\theta)}{\partial \theta}\bigg)^{\T}  \ve{R}_{\ve{\phi}}^{-1}(\theta) \big( \ve{\tilde{\phi}}-\ve{\mu}_{\ve{\phi}}(\theta)\big)=0
\end{align}
for $\theta$, results in a consistent estimate $\hat{\theta}(\ve{Z})$ with asymptotic variance equal to the inverse of $N$ times the right-hand side of the Fisher information bound \eqref{bound:eigenvalue} \cite{SteinTSP16}.
\section{$1$-bit DOA Estimation - System Model}
For the application of these results to blind DOA parameter estimation with $1$-bit ADC, we assume a uniform linear array (ULA) with $K$ sensors, where the spacing between the antennas is equal to half the wavelength. With a signal source 
\begin{align}
\ve{x}=\begin{bmatrix}x_{I} &x_{Q} \end{bmatrix}^{\T}\in\fieldR^2
\end{align}
consisting of independent zero-mean Gaussian in-phase and quadrature components with unit covariance
\begin{align}
\exdi{\ve{x}}{\ve{x}\ve{x}^{\T}}&=\ve{I}_2
\end{align}
and under a narrowband assumption, the unquantized receive signal of size $M=2K$
\begin{align}
\ve{y}=\begin{bmatrix}\ve{y}^{\T}_{I} &\ve{y}^{\T}_{Q} \end{bmatrix}^{\T}\in\fieldR^{M},\label{eq:system_model}
\end{align}
can be written in a real-valued notation \cite{SteinWSA13}
\begin{align}
\ve{y}=\gamma\ve{A}(\theta)\ve{x}+\ve{\eta},
\end{align}
where $\theta$ is the direction under which the transmit signal $\ve{x}$ impinges on the receive array. Note, that $\ve{\eta}\in\fieldR^{M}$ is independent zero-mean additive Gaussian noise with unit variance
\begin{align}
\exdi{\ve{\eta}}{\ve{\eta}\ve{\eta}^{\T}}=\ve{I}_{M}.
\end{align}
The full array steering matrix \cite{SteinWSA13}
\begin{align}
\ve{A}(\theta)=\begin{bmatrix}\ve{A}^{\T}_I(\theta) &\ve{A}^{\T}_Q(\theta) \end{bmatrix}^{\T}\in\fieldR^{M \times 2},
\end{align}
is modulated by the DOA parameter $\theta\in\fieldR$ and consists of an in-phase steering matrix
\begin{align}
\ve{A}_I(\theta)=\begin{bmatrix}
\xi_1(\theta) &\psi_1(\theta)\\ 
\xi_2(\theta) &\psi_2(\theta) \\ 
\vdots &\vdots\\ 
\xi_K(\theta) &\psi_K(\theta)
\end{bmatrix}\in\fieldR^{K \times 2}
\end{align}
and a quadrature steering matrix
\begin{align}
\ve{A}_Q(\theta)=\begin{bmatrix}
-\psi_1(\theta) &\xi_1(\theta)\\ 
-\psi_2(\theta) &\xi_2(\theta) \\ 
\vdots &\vdots\\ 
-\psi_K(\theta) &\xi_K(\theta)
\end{bmatrix}\in\fieldR^{K \times 2},
\end{align}
with entries
\begin{align}
\xi_k(\theta)&=\cos{\big((k-1)\pi\sin{(\theta)}\big)}\notag\\
\psi_k(\theta)&=\sin{\big((k-1)\pi\sin{(\theta)}\big)}.
\end{align}
Therefore, the parametric covariance of the receive signal is
\begin{align}
\exdi{\ve{y};\theta}{\ve{y}\ve{y}^{\T}}&=\ve{\Sigma}_{y}(\theta)\notag\\
&={\gamma^2}\ve{A}(\theta)\ve{A}^{\T}(\theta)+\ve{I}_{2K}
\label{eq:receive:covariance}
\end{align}
and the $1$-bit receive signal can be modeled 
\begin{align}
\ve{z}=\sign{\ve{y}}
\end{align}
as discussed in section \ref{sec:motivation}.
\section{$1$-bit DOA Estimation - Performance Analysis}
\subsection{$1$-bit Exponential Replacement}
In order to apply the pessimistic approximation of the Fisher information \eqref{bound:eigenvalue} for DOA estimation after $1$-bit hard-limiting, we use the auxiliary statistics
\begin{align}
\ve{\phi}(\ve{z})&=\vech{\ve{z} \ve{z}^{\T}},
\end{align}
where $\vech{\ve{B}}$ denotes the half-vectorization of the symmetric matrix $\ve{B}$, i.e., the vectorization of the lower triangular part of $\ve{B}$. The required mean \eqref{aux:mean} is given by
\begin{align}
\ve{\mu}_{\ve{\phi}}(\theta)&=\exdi{\ve{z};\theta}{\ve{\phi}(\ve{z})}\notag\\
&=\exdi{\ve{z};\theta}{\vech{\ve{z} \ve{z}^{\T}}}\notag\\
&=\vech{\exdi{\ve{z};\theta}{\ve{z} \ve{z}^{\T}}}\notag\\
&= \vech{ \ve{\Sigma}_{z}(\theta) },
\end{align}
where by the arcsine law \cite[pp. 284]{Thomas69}
\begin{align}
\ve{\Sigma}_{z}(\theta)=\frac{2}{\pi} \arcsin{ \frac{1}{\gamma^2 + 1} \ve{\Sigma}_{y}(\theta) }.
\end{align}
For the derivative of the mean we find
\begin{align}
\frac{\partial \ve{\mu}_{\ve{\phi}}(\theta)}{\partial \theta} = \vech{ \frac{\partial \ve{\Sigma}_{z}(\theta)}{\partial \theta} },
\end{align}
where the derivative of the quantized covariance matrix is
\begin{align}
&\left[ \frac{\partial \ve{\Sigma}_{z}(\theta)}{\partial \theta} \right]_{ij} =  \frac{2 \left[ \frac{\partial \ve{\Sigma}_{y}(\theta)}{\partial \theta} \right]_{ij}}{\pi(\gamma^2+1)\sqrt{1-\frac{1}{(\gamma^2+1)^2}  \left[\ve{\Sigma}_{y}(\theta)\right]_{ij}^2}}
\end{align}
with the derivative of the unquantized covariance being
\begin{align}
\frac{\partial \ve{\Sigma}_{y}(\theta)}{\partial \theta}=\gamma^2 \Bigg(\frac{ \partial \ve{A}(\theta)}{\partial \theta}\ve{A}^{\T}(\theta)+\ve{A}(\theta)\frac{ \partial \ve{A}^{\T}(\theta)}{\partial \theta}\Bigg).
\end{align}
The derivative of the steering matrix is
\begin{align}
\frac{\partial \ve{A}(\theta)}{\partial \theta}=\begin{bmatrix}\frac{\partial \ve{A}^{\T}_I(\theta)}{\partial \theta} &\frac{\partial \ve{A}^{\T}_Q(\theta)}{\partial \theta} \end{bmatrix}^{\T},
\end{align}
with the in-phase component
\begin{align}
\frac{\partial \ve{A}_I(\theta)}{\partial \theta}=\begin{bmatrix}
\frac{\partial \xi_1(\theta)}{\partial \theta} &\frac{\partial \psi_1(\theta)}{\partial \theta}\\ 
\frac{\partial \xi_2(\theta)}{\partial \theta} &\frac{\partial \psi_2(\theta)}{\partial \theta} \\ 
\vdots &\vdots\\ 
\frac{\partial \xi_K(\theta)}{\partial \theta} &\frac{\partial \psi_K(\theta)}{\partial \theta}
\end{bmatrix}
\end{align}
and a quadrature component
\begin{align}
\frac{\partial \ve{A}_Q(\theta)}{\partial \theta}=\begin{bmatrix}
-\frac{\partial \psi_1(\theta)}{\partial \theta} &\frac{\partial \xi_1(\theta)}{\partial \theta}\\ 
-\frac{\partial \psi_2(\theta)}{\partial \theta} &\frac{\partial \xi_2(\theta)}{\partial \theta}\\ 
\vdots &\vdots\\ 
-\frac{\partial \psi_K(\theta)}{\partial \theta} &\frac{\partial \xi_K(\theta)}{\partial \theta}
\end{bmatrix},
\end{align}
while the individual entries are
\begin{align}
\frac{\partial \xi_k(\theta)}{\partial \theta}&=-(k-1)\pi\cos{(\theta)}\sin{\big((k-1)\pi\sin{(\theta)}\big)}\notag\\
\frac{\partial \psi_k(\theta)}{\partial \theta}&=(k-1)\pi\cos{(\theta)}\cos{\big((k-1)\pi\sin{(\theta)}\big)}.
\end{align}
For the second moment of the auxiliary statistics \eqref{aux:covariance}
\begin{align}
\exdi{\ve{z};\theta}{\ve{\phi}(\ve{z}) \ve{\phi}^{\T}(\ve{z})}=\exdi{\ve{z};\theta}{\vech{\ve{z} \ve{z}^{\T}}\vech{\ve{z} \ve{z}^{\T}}^{\T}}
\end{align}
is required. This implies to evaluate the expected value
\begin{align}
\ex{z_i z_j z_k z_l},\quad\quad i, j, k, l \in \{1,\ldots, M\}.
\end{align}
For the cases $i=j=k=l$ or $i=j \neq k=l$, we obtain
\begin{align}
\ex{z_i z_j z_k z_l}&=\ex{z^4_i}\notag\\
&=\ex{z^2_i z^2_k}\notag\\
&=1.
\end{align}
If $i=j=k \neq l$, the arcsine law results in
\begin{align}
\ex{z_i z_j z_k z_l}&=\ex{z^3_i z_l}\notag\\
&=\ex{z_i z_l}\notag\\
&=\frac{2}{\pi} \arcsin{\rho_{il}(\theta)},
\end{align}
like in the case $i=j\neq k \neq l$, where
\begin{align}
\ex{z_i z_j z_k z_l}&=\ex{z^2_i z_k z_l}\notag\\
&=\ex{z_k z_l}\notag\\
&=\frac{2}{\pi} \arcsin{\rho_{kl}(\theta)}.
\end{align}
The case  $i \neq j \neq k \neq l$ requires special care, as
\begin{align}
\ex{z_i z_j z_k z_l}&=\Prob{z_i z_j z_k z_l=1}\notag\\
&\quad-\Prob{z_i z_j z_k z_l=-1}
\end{align}
involves the evaluation of the $2^4=16$ orthant probabilities 
\begin{align}
\Phi_q(\ve{\bar{\Sigma}}(\theta))=\Prob{\pm z_i >0, \pm z_j>0, \pm z_k>0, \pm z_l>0}
\end{align}
of a quadrivariate Gaussian variable with correlation matrix
\begin{align}
\ve{\bar{\Sigma}}(\theta)=
\begin{bmatrix} 
1 &\rho_{ij}(\theta) &\rho_{ik}(\theta) &\rho_{il}(\theta)\\
\rho_{ij}(\theta) &1&\rho_{jk}(\theta) &\rho_{jl}(\theta)\\
\rho_{ik}(\theta) &\rho_{jk}(\theta) &1 &\rho_{kl}(\theta)\\
\rho_{il}(\theta) &\rho_{jl}(\theta)&\rho_{kl}(\theta) &1\\
\end{bmatrix}.
\end{align}
A closed-form solution for this problem, requiring calculation of four one-dimensional integrals, is given in \cite{Sinn11}. 
\subsection{$1$-bit Quantization Loss}
Using result \eqref{bound:eigenvalue}, we can evaluate the quantization loss for DOA parameter estimation for receivers with $K>2$ in a pessimistic manner by forming the ratio
\begin{align}\label{def:qloss}
\chi(\theta)=\frac{\big(\frac{\partial \ve{\mu}_{\ve{\phi}}(\theta)}{\partial \theta}\big)^{\T}  \ve{R}_{\ve{\phi}}^{-1}(\theta)\frac{\partial \ve{\mu}_{\ve{\phi}}(\theta)}{\partial \theta}}{F_{\ve{y}}(\theta)}.
\end{align}
In Fig. \ref{QLoss_SNR} we plot the performance loss \eqref{def:qloss} versus the signal-to-noise ratio 
\begin{align}
\text{SNR}={\gamma^2} 
\end{align}
for two different DOA setups ($\theta=10^\circ$ and $\theta=70^\circ$).
\pgfplotsset{legend style={rounded corners=2pt,nodes=right}}
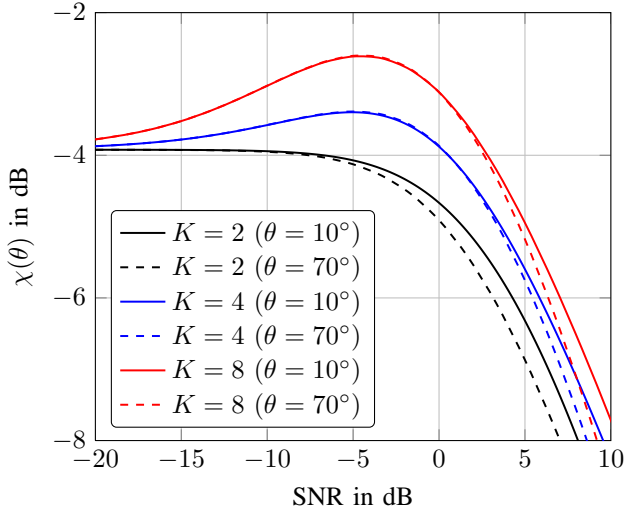
\begin{figure}[!htbp]
\begin{tikzpicture}[scale=1.0]

  	\begin{axis}[ylabel=$\chi(\theta)\text{ in dB}$,
  			xlabel=$\text{SNR in dB}$,
			grid,
			ymin=-8,
			ymax=-2,
			xmin=-20,
			xmax=10,
			legend pos=south west]
			
    			\addplot[black, style=solid, line width=0.75pt,smooth] table[x index=0, y index=1]{LossSNR_K2_zeta10.txt};
			\addlegendentry{$K=2$ $(\theta=10^\circ)$}
			
			\addplot[black, style=dashed, line width=0.75pt,smooth] table[x index=0, y index=1]{LossSNR_K2_zeta70.txt};
			\addlegendentry{$K=2$ $(\theta=70^\circ)$}
			
    			\addplot[blue, style=solid, line width=0.75pt,smooth] table[x index=0, y index=1]{LossSNR_K4_zeta10.txt};
			\addlegendentry{$K=4$ $(\theta=10^\circ)$}
				
			\addplot[blue, style=dashed, line width=0.75pt,smooth] table[x index=0, y index=1]{LossSNR_K4_zeta70.txt};
			\addlegendentry{$K=4$ $(\theta=70^\circ)$}
			
			\addplot[red, style=solid, line width=0.75pt,smooth] table[x index=0, y index=1]{LossSNR_K8_zeta10.txt};	
			\addlegendentry{$K=8$ $(\theta=10^\circ)$}

			\addplot[red, style=dashed, line width=0.75pt,smooth] table[x index=0, y index=1]{LossSNR_K8_zeta70.txt};	
			\addlegendentry{$K=8$ $(\theta=70^\circ)$}
			
\end{axis}	
\end{tikzpicture}
\caption{$1$-bit DOA Estimation - Quantization Loss vs. SNR}
\label{QLoss_SNR}
\end{figure}
It can be observed that the quantization loss $\chi(\theta)$ becomes smaller for arrays with a larger number of antennas $K$. In particular, the array size plays a beneficial role in the SNR range of $-10$ to $0$ dB, which is a regime of high practical relevance for energy-efficient broadband mobile communication systems. However, for situations where $\text{SNR}>5$ dB the quantization loss becomes pronounced. In Fig. \ref{QLoss_Angle} the performance loss is depicted as a function of the DOA parameter $\theta$ for three different array sizes ($K=2,4,8$).
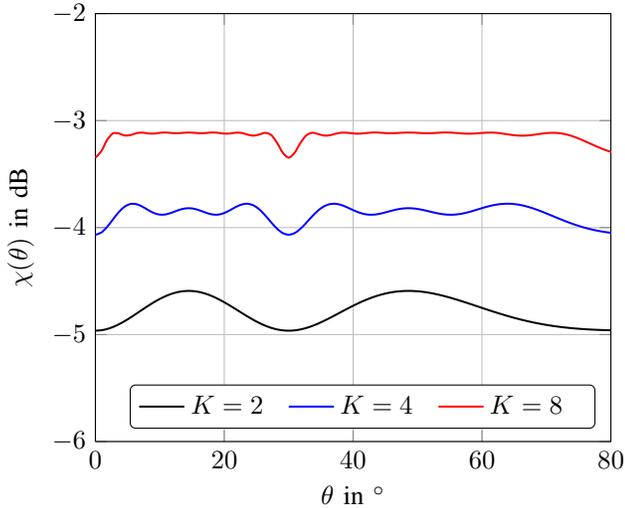
\begin{figure}[!htbp]
\begin{tikzpicture}[scale=1.0]

  	\begin{axis}[ylabel=$\chi(\theta)\text{ in dB}$,
  			xlabel=$\theta\text{ in $^\circ$}$,
			grid,
			ymin=-6,
			ymax=-2,
			xmin=0,
			xmax=80,
			legend columns=-1,
			legend pos=south east]
			
    			\addplot[black, style=solid, line width=0.75pt,smooth] table[x index=0, y index=1]{LossAngle_K2_SNR0.txt};
			\addlegendentry{$K=2\text{\phantom{X}}$}
			
    			\addplot[blue, style=solid, line width=0.75pt,smooth] table[x index=0, y index=1]{LossAngle_K4_SNR0.txt};
			\addlegendentry{$K=4\text{\phantom{X}}$}
				
			\addplot[red, style=solid, line width=0.75pt,smooth] table[x index=0, y index=1]{LossAngle_K8_SNR0.txt};	
			\addlegendentry{$K=8\text{\phantom{X}}$}
			
\end{axis}	
\end{tikzpicture}
\caption{$1$-bit DOA Estimation - Quantization Loss vs. DOA (SNR$=0$ dB)}
\label{QLoss_Angle}
\end{figure}
It is visible that the quantization loss $\chi(\theta)$ becomes less dependent on the DOA parameter $\theta$ for large arrays. Finally, in Fig. \ref{QLoss_K} the quantization loss $\chi(\theta)$ is shown for a growing number of antennas $K$. For a low SNR setup ($\text{SNR}=-15$ dB) the gap between the quantized receiver and the unquantized receiver vanishes linearly with the array size $K$, while for a medium SNR scenario ($\text{SNR}=-3$ dB) the relative performance of the $1$-bit receive system strongly improves by increasing the amount of receive sensors $K$.
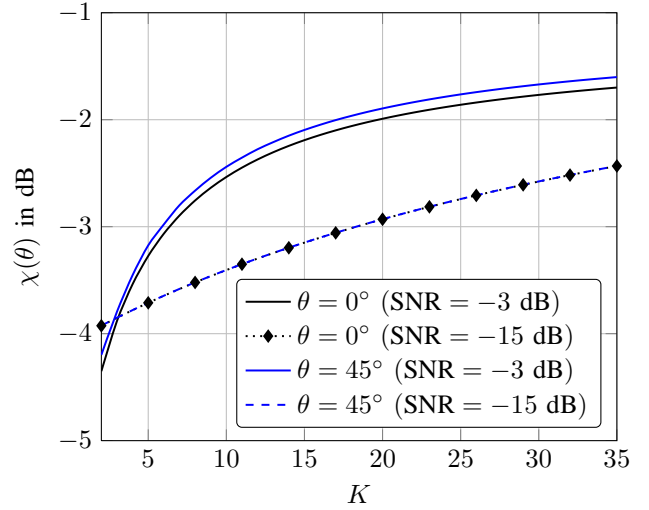
\begin{figure}[!htbp]
\begin{tikzpicture}[scale=1.0]

  	\begin{axis}[ylabel=$\chi(\theta)\text{ in dB}$,
  			xlabel=$K$,
			grid,
			ymin=-5,
			ymax=-1,
			xmin=2,
			xmax=35,
			legend pos=south east]
			
    			\addplot[black, style=solid, line width=0.75pt,smooth] table[x index=0, y index=1]{LossK_SNR-3_zeta0.txt};
			\addlegendentry{$\theta=0^\circ$ $(\text{SNR}=-3\text{ dB})$}
			
			\addplot[black, style=dotted, line width=0.75pt,smooth, every mark/.append style={solid}, mark=diamond*, mark repeat=3] table[x index=0, y index=1]{LossK_SNR-15_zeta0.txt};
			\addlegendentry{$\theta=0^\circ$ $(\text{SNR}=-15\text{ dB})$}
			
    			\addplot[blue, style=solid, line width=0.75pt,smooth] table[x index=0, y index=1]{LossK_SNR-3_zeta45.txt};
			\addlegendentry{$\theta=45^\circ$ $(\text{SNR}=-3\text{ dB})$}
			
			\addplot[blue, style=dashed, line width=0.75pt,smooth] table[x index=0, y index=1]{LossK_SNR-15_zeta45.txt};
			\addlegendentry{$\theta=45^\circ$ $(\text{SNR}=-15\text{ dB})$}
				
			
			
\end{axis}	
\end{tikzpicture}
\caption{$1$-bit DOA Estimation - Quantization Loss vs. Array Elements}
\label{QLoss_K}
\end{figure}
\subsection{$1$-bit CMLE Algorithm}
\begin{figure}[!htbp]
\begin{tikzpicture}[scale=1.0]

  	\begin{axis}[ylabel=$\text{RMSE in }^\circ$,
  			xlabel=$\text{SNR in dB}$,
			grid,
			ymin=0.005,
			ymax=0.015,
			xmin=-6,
			xmax=0,
			legend pos=north east]
			
			\addplot[red, style=dashed, line width=0.75pt,mark=*] table[x index=0, y index=2]{DOA1bit_MLE_K4_zeta5.txt};
			\addlegendentry{RMSE}
			
			
			\addplot[black, style=solid, line width=0.75pt] table[x index=0, y index=1]{DOA1bit_MLE_K4_zeta5.txt};
			\addlegendentry{PCRLB}
			
										
\end{axis}	
\end{tikzpicture}
\caption{$1$-bit DOA Estimation - CMLE Performance ($K=4$, $\theta=5^\circ$)}
\label{CMLE_Performance}
\end{figure}
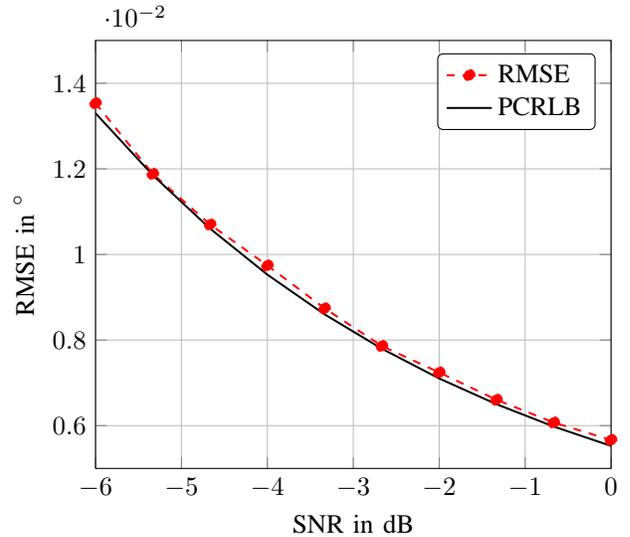

In order to demonstrate that the framework of exponential replacement also provides a useful guideline how to achieve the guaranteed performance
\begin{align}
\exdi{\ve{Z};\theta}{\big(\theta-\hat{\theta}(\ve{Z})\big)^{2}}& \approx \frac{1}{N\Big(\frac{\partial \ve{\mu}_{\ve{\phi}}(\theta)}{\partial \theta}\Big)^{\T}  \ve{R}_{\ve{\phi}}^{-1}(\theta)\frac{\partial \ve{\mu}_{\ve{\phi}}(\theta)}{\partial \theta}}\notag\\
&=\text{PCRLB},
\end{align}
in Fig. \ref{CMLE_Performance} we plot the accuracy (RMSE) of the conservative maximum-likelihood estimation (CMLE) algorithm \eqref{algo:cmle} for an array size of $K=4$, a DOA parameter $\theta=5^\circ$ and $N=1000$ samples averaged over $10000$ runs.
It can be observed that the CMLE performs close to the pessimistic version of the CRLB.
\section{Conclusion}
We have discussed the method of exponential replacement \cite{SteinTSP16} in the context of $1$-bit DOA estimation with a single signal source and a receive array of $K$ sensors. The associated pessimistic approximation for the Fisher information measure allows to analyze the achievable DOA estimation accuracy for arrays with $K>2$ in a conservative manner. Additionally, the framework provides a guideline how to achieve the accuracy guaranteed by the pessimistic CRLB. The performance analysis shows that in the medium SNR regime DOA estimation with $1$-bit ADC can be performed at high accuracy if the number of array elements $K$ is large. This result supports the current discussion on future wireless systems which use a large number of low-complexity sensors, i.e., $1$-bit massive MIMO communication systems \cite{Risi14} \cite{Jacobsson15}.
\bibliographystyle{IEEEbib}

\end{document}